# Blind Source Separation of Optical Wireless Communications in Noisy Environments

**Pengfei Xu[a,*]**, **Yinjie Jia[a,b]** and **Zhijian Wang[a]**

[a] College of Computer and Information, Hohai University, Nanjing, Jiangsu 210098, China;
[b] Faculty of Electronic Information Engineering, Huaiyin Institute of Technology, Huaian, Jiangsu 223003, China;

**Abstract:** Blind source separation is a research hotspot in the field of signal processing because it aims to separate unknown source signals from observed mixtures through an unknown transmission channel. A low computational complexity instantaneous linear mixture signals blind separation algorithm was introduced and improved. There is only one variable parameter named the length of moving average in the algorithm, which has a significant impact on the separation effect. This paper gives some suggestions on the reasonable value through experiments. The algorithm is extended to the separation of visible light communication signals in different noise environments, and has achieved certain results, which further expands the applicability of the algorithm.

## 1. Introduction

The seminal work on blind source separation(BSS) is by Jutten and Herault [1] in 1985, the problem is to extract the underlying source signals from a set of mixtures, where the mixing matrix is unknown. In other words, BSS seeks to recover original source signals from their mixtures without any prior information on the sources or the parameters of the mixtures. Its research results have been widely applied in many fields, such as speech recognition, optical wireless communication[2-5], biomedicine, image processing and so on.

Nowadays, there have been lots of blind source separation algorithms to calculate a demixing matrix , so we can make the estimated source signal only by the received signal. In this paper we select the blind source separation algorithm based on maximum SNR [6]. Some literatures have improved and optimized this algorithm. blind source separation based on maximum SNR is adopted to calculate the unmixing matrix in cooperative spectrum sensing[7]. The negative entropy is used to multiply the SNR and taken as a new independence measurement function to improve the BSS algorithm[8].

Compared to other blind source separation algorithms, demixing matrix can be achieved without any iterative, the merit of selected algorithm is very low computational complexity. Therefore, blind source separation based on maximum SNR is adopted to calculate the demixing matrix in this paper.

The rest of the paper is organized as follows. In Section 2, we introduce the noisy signal BSS model and maximum SNR algorithm. In Section 3, the simulation experiment that indicates the effectiveness of this optimized method is presented. The final section is a summary of the content of this paper and some questions that need further study.

## 2. Methodology

*2.1. Noisy Signal BSS Model*

Source signals $s_i(t)$ come from different signal sources (assumes that the signal is continuous signal ), so $s_i(t)$ can be think mutual statistical independence, $x_i(t)$ is mixed signals or observation signals(See Fig.1).

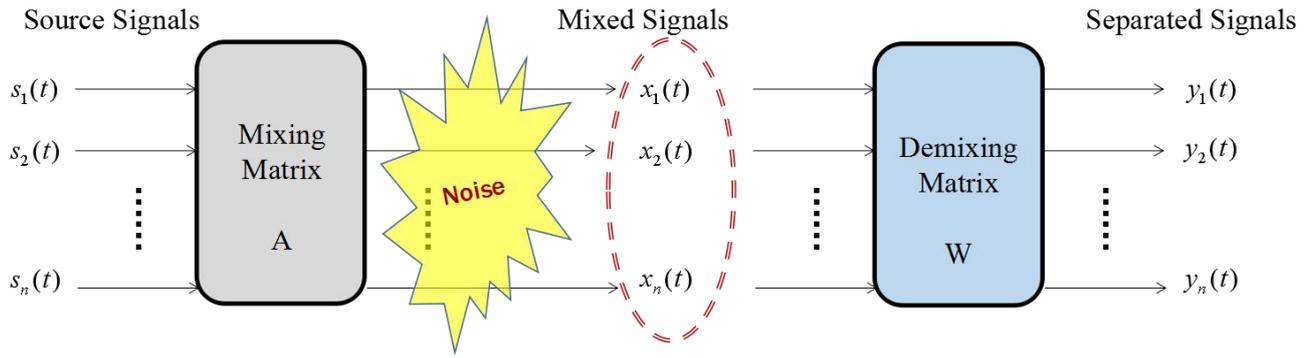

**Fig. 1.** Noisy signal blind source separation model

The problem of basic linear BSS can be expressed algebraically as follows:

$$x_i(t) = \sum_1^n a_{ij}(s_i(t) + v(t)) \tag{1}$$

Where $a_{ij}$ is mixed coefficient, formula (1) can be write in vector as follow:

$$x(t) = A(s(t) + v(t)) \tag{2}$$

Where $s(t) = [s_1(t) \cdots s_n(t)]^T$ is a column vector of Source signals, $x(t) = [x_1(t) \cdots x_n(t)]^T$ is vector of mixed signals or observation signals, $v(t)$ is additive white gaussian noise, $A$ is $n \times n$ mixing matrix. Problem of blind source separation (BSS) only know observation signals and statistical independence property of Source signals. In virtue of the knowledge of probability distribution of Source signals we can recover Source signals. Assume $W$ is $n \times n$ demixing matrix or separating matrix, problem of BSS can be describe as follow:

$$y(t) = Wx(t) \tag{3}$$

Where $y(t)$ is a estimate of or separated signals. To compare Recovered signals $y(t)$ with Source signals $s(t)$, There are changes in magnitude and order, which is called the inherent uncertainty or indeterminations of blind source separation. Because of signal information include in waveform, so indeterminations of permutation and scale do not influence applications of BSS.

BSS has two steps, first create a cost function $F(W)$ with respect to $W$, if $\hat{W}$ can make $F(W)$ reach to maximum, $\hat{W}$ is the demixing matrix. Second is to find a effective iterative algorithm for solution of $\frac{\partial F}{\partial W} = 0$. In this paper, cost function is the function of signal noise ratio (SNR), optimize processing of cost function result in generalized eigenvalue problem, demixing matrix was achieved by solving the generalized eigenvalue problem without any iterative. The BSS in the paper has high calculation efficiency and simulation result is better.

*2.2. Signal-to-noise ratio objective function*

Maximum signal-to-noise ratio (MSNR) algorithm belongs to matrix eigenvalue decomposition method. By constructing the signal-to-noise ratio contrast function and estimating the separation matrix by eigenvalue decomposition or generalized eigenvalue decomposition, the closed-form solution can be found directly without iterative optimization process. Therefore, it has the advantages of simple algorithm and fast running speed, and is convenient for real-time processing and hardware implementation of FPGA. The time continuous radio signal is sampled and changed into a discrete value. In the following formula, the time mark $t$ becomes $n$.

According to the model of blind source separation, the error $e(n) = s(n) - y(n)$ between the source signal $s(n)$ and the output signal $y(n)$ is regarded as noise. When the minimum value of $e(n)$ is taken, the estimated value $y(n)$ is the optimal approximation of the source signal $s(n)$, and the effect of blind source separation is the best. The power ratio of source signal $s(n)$ to $e(n)$ is defined as signal-to-noise ratio. When $e(n)$ is the smallest, it is equivalent to the largest signal-to-noise ratio. According to this estimation criterion, the signal-to-noise ratio function [2] is constructed as follow:

$$F_{SNR} = 10\log\frac{s \cdot s^T}{e \cdot e^T} = 10\log\frac{s \cdot s^T}{(s-y) \cdot (s-y)^T} \qquad (4)$$

Because the source signal $s(n)$ is unknown, the mean value of noise is 0, so we use moving average of estimate signals $\bar{y}(n)$ instead of source signals $s(n)$. Formula (4) can be write as:

$$F_{SNR} = 10\log\frac{s \cdot s^T}{e \cdot e^T} = 10\log\frac{\bar{y} \cdot \bar{y}^T}{(\bar{y}-y) \cdot (\bar{y}-y)^T} \qquad (5)$$

Where $\bar{y}_i(n) = \frac{1}{L}\sum_{j=0}^{L} y_i(n-j), j = 0,1,2,....,L-1$ is moving average of estimate signals $y(n)$. Integer $L$ is the length of moving average. In order to simplify calculation, we replace $\bar{y}(n)$ with $y(n)$ in the molecule of formula (5). So we gained maximum signal noise ratio cost function as follow:

$$F_{SNR}* = 10\log\frac{y \cdot y^T}{(\bar{y}-y) \cdot (\bar{y}-y)^T} \qquad (6)$$

According to formula (3), we get the formula (7) as follows.
$$\bar{y}(n) = W\bar{x}(n) \qquad (7)$$

Where $\bar{x}(n) = \frac{1}{L}\sum_{j=0}^{L} x_i(n-j), j = 0,1,2,.....,L-1$ is a moving average of mixed signals $x(n)$.

We substitute formula (3) and formula (7) into formula (6) and Formula (8) is deduced.

$$F_{SNR}*(W) = 10\log\frac{y \cdot y^T}{(\bar{y}-y) \cdot (\bar{y}-y)^T} = 10\log\frac{Wx \cdot x^TW^T}{W(\bar{x}-x) \cdot (\bar{x}-x)^TW^T}$$
$$= 10\log\frac{WCW^T}{W\bar{C}W^T} = 10\log\frac{V}{U} \qquad (8)$$

Where $\bar{C} = (\bar{x}-x)(\bar{x}-x)^T$ and $C = xx^T$ are correlation matrixs, $U = W\bar{C}W^T; V = WCW^T$.

*2.3. Derivation of Separation Algorithms*

According to formula (8), derivative of F with respect to is:

$$\frac{\partial F_{SNR}*(W)}{\partial W} = \frac{2W}{V}C - \frac{2W}{U}\bar{C} \qquad (9)$$

According to the definition, when the maximum value of the function $F_{SNR}*(W)$ is obtained, the gradient is 0.

So we get the following formula.

$$WC = \frac{V}{U}W\overline{C} \quad (10)$$

We can obtain demixing matrix $\hat{W}$ by solving formula (10), it has been proved solution of formula (10) that is eigenvector of $\tilde{C} \cdot C^{-1}$ in literature [9]. All Source signals can be recovered once: $y = \hat{W}x$, where each row of y corresponds to exactly one extracted signal $y_i$.

## 3. Experiments and Results

A simple model of optical wireless communications is shown in Figure 2. First, the binary input signal is optically modulated (QPSK and OOK are used here) and emitted by the light emitting diode. Then it pass through the indoor channel and detected by the receiver's photoelectric detector. Finally, the mixed signal $x(n)$ is obtained. In this section, the mixed signals $x(n)$ is separated by the blind signal processing algorithm proposed above.

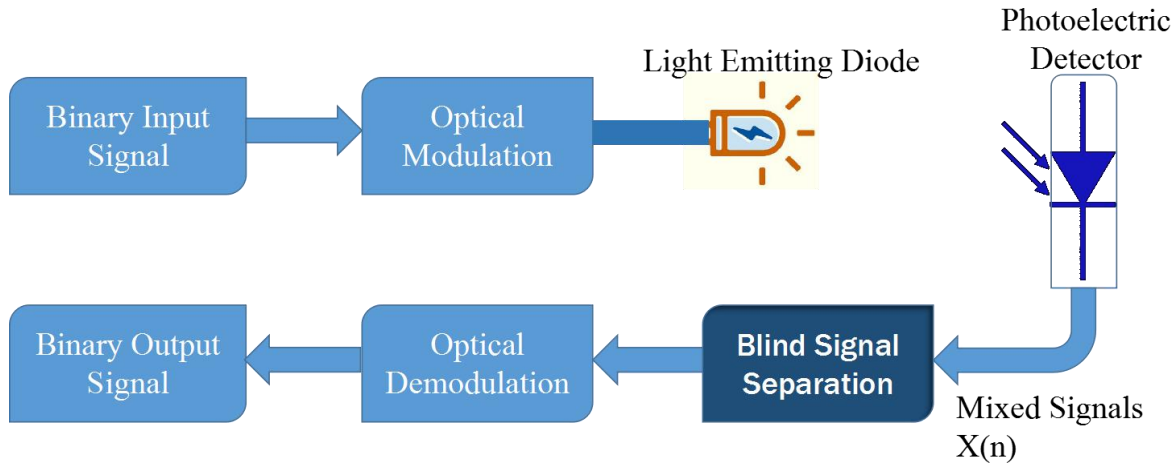

**Fig. 2.** Blind source separation of optical wireless communications

In this experiment(MATLAB R2017b), we randomly generate two sets of binary input signals Bit1 and Bit2. They take values such as Bit1= '0 0 0 1 1 0 1 0 0 1 0 1 1 1 1 0 1 0 0 1' and Bit2= '0 0 0 0 0 0 0 0 1 1 0 0 1 1 0 0 0 1 0 1'. Bit1 uses QPSK modulation as the source signal S1, Bit2 uses OOK modulation as the source signal S2. The AWGN (Additive White Gaussian Noise) channel model is used for noise and interference in optical signal transmission channel. Its amplitude distribution obeys the Gauss distribution, and the power spectral density is uniformly distributed. The signal-to-noise ratio(SNR) is 30dB here. The mixing matrix $A$ was randomly generated(such as $A$ =[0.4684 0.1952; 0.7384 0.5483]). The number of samples N=2000, and the length of moving average $L$ =7(See Fig. 3).

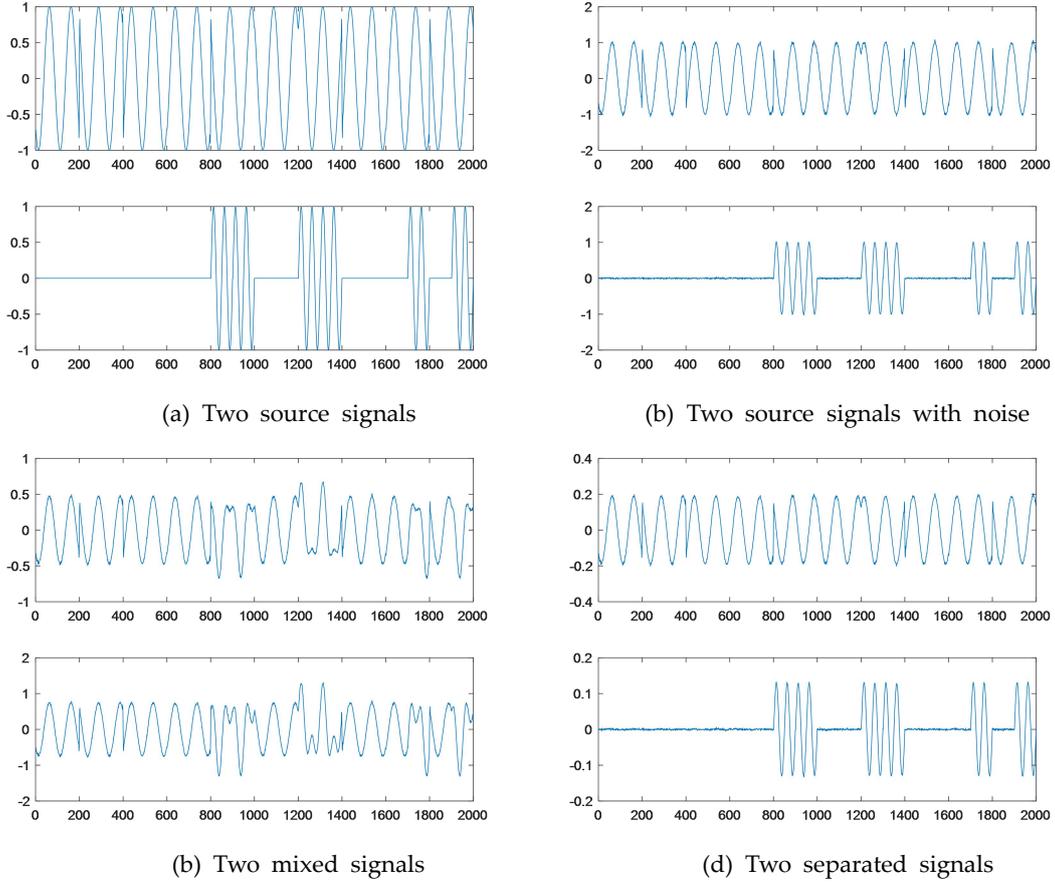

(a) Two source signals　　　　　(b) Two source signals with noise

(b) Two mixed signals　　　　　(d) Two separated signals

**Fig. 3.** Separation of mixed signals

The mixed signals have been successfully separated. We can further denoise the separated signals [10] and improve the performance of optical wireless communication system.

Evaluating the performance of blind source separation, a correlation coefficient $C$ is introduced as a performance index.

$$C(x,y) = \frac{\text{cov}(x,y)}{\sqrt{\text{cov}(x,x)}\sqrt{\text{cov}(y,y)}} \qquad (11)$$

$C(x,y) = 0$ means that x and y are uncorrelated, and the signals correlation increases as $C(x,y)$ approaches unity, the signals become fully correlated as $C(x,y)$ becomes unity. The value of $L$ can choose integers less than 100 [2], this value is relatively broad. It has a significant impact on the separation effect. In previous literatures, its role is often neglected, but only based on experience. Therefore,we extend the experimental conditions based on this simulation, the separation effect is analyzed in different sliding window lengths and different noise here(see Fig.4).

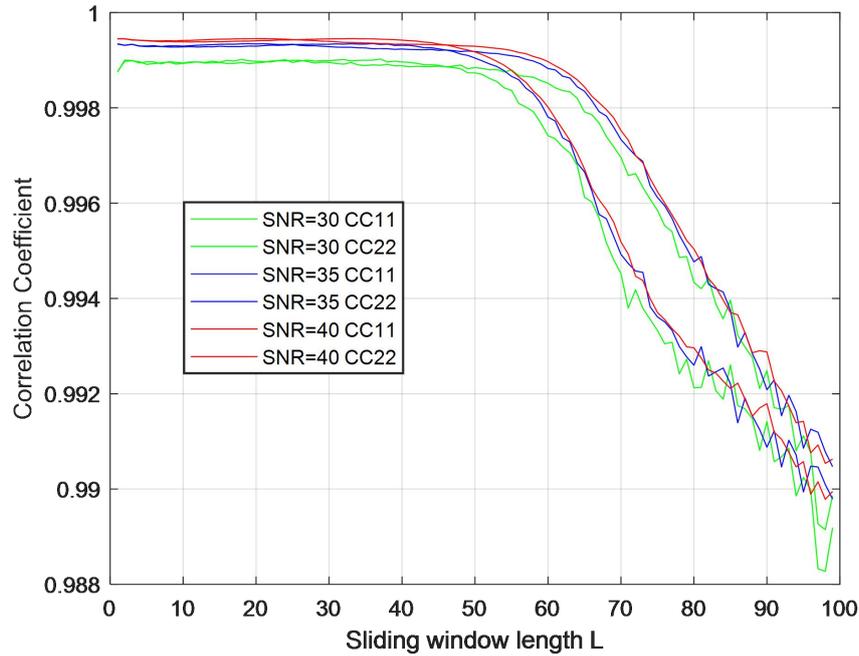

**Fig.4.** Effect under different sliding window lengths and SNR

It is known from the figure that: (1) in different noise environments, the less the noise, the better the separation effect of the algorithm. (2) The optimum value of sliding window length $L$ is about [2,30]. After the sliding window length exceeds 30, the separation effect begins to deteriorate.

## 4. Conclusions

In this paper, the maximum signal-to-noise ratio blind source separation algorithm is used for reference and optimized. We looks into the possibility of its application in optical wireless communications. Firstly, the algorithm is introduced and applied to the separation of optical wireless communication signals in noisy environment, which further expands the applicability of the algorithm. Then the optimal range of sliding average window in the algorithm is discussed by experimental analysis, and the algorithm is optimized. Due to its simple principle and good transplantation capability, it can be widely applied in various fields of digital signal processing.